\documentclass[prl,aps,preprint,superscriptaddress]{revtex4-1}  

\usepackage{graphicx} 
\usepackage{siunitx}

\begin{document}

\title{Magnetotactic bacteria  in a droplet self-assemble into a rotary motor}

\author{Benoit Vincenti}
\affiliation{Laboratoire PMMH, UMR 7636 CNRS-ESPCI-Sorbonne Universit\'e-Universit\'e Paris Diderot, 7-9 quai Saint-Bernard, 75005 Paris, France.}

\author{Gabriel Ramos, Maria Luisa Cordero}
\affiliation{Departamento de F\'\i sica, FCFM, Universidad de Chile, Av.\ Blanco Encalada 2008, Santiago, Chile.}

\author{Carine Douarche}
\affiliation{Laboratoire FAST, Univ. Paris-Sud, CNRS, Universit\'e Paris-Saclay, F-91405, Orsay, France.}

\author{Rodrigo Soto}
\affiliation{Departamento de F\'\i sica, FCFM, Universidad de Chile, Av.\ Blanco Encalada 2008, Santiago, Chile.}

\author{Eric Clement}
\affiliation{Laboratoire PMMH, UMR 7636 CNRS-ESPCI-Sorbonne Universit\'e-Universit\'e Paris Diderot, 7-9 quai Saint-Bernard, 75005 Paris, France.}

\begin{abstract}
From intracellular protein trafficking to large scale motion of animal groups, the physical concepts driving the self-organization of living systems are still largely unraveled. Self-organization of active entities, leading to novel phases and emergent macroscopic properties, recently shed new lights on these complex dynamical processes. Here we show that, under the application of a constant magnetic field, motile magnetotactic bacteria confined in water-in-oil droplets self-assemble into a rotary motor exerting a torque on the external oil phase. A collective motion in the form of a large-scale vortex, reversable by inverting the field direction, builds-up in the droplet with a vorticity perpendicular to the magnetic field. We study this collective organization at different concentrations, magnetic fields and droplets radii and reveal the formation of two torque-generating areas close to the droplet interface. We characterize quantitatively the mechanical energy extractable from this new biological and self-assembled motor. 
\end{abstract}

\maketitle

\section*{Introduction}
One of the hallmark of life in its most phenomenological appearance is its ability to spontaneously produce collective, hierarchical and organized motion. However, it has been shown that non-living entities can also develop complex and coordinated structures when driven out of thermodynamic equilibrium~\cite{Whitesides2002}. So far, a complete picture guiding the emergence of these active patterns is still lacking. Indeed, since the pioneering work of Vicseck~\cite{Vicsek2012}, physics communities have constantly tried to unravel the basic principles leading to collective pattern formation of active particles~\cite{Marchetti2013,Bechinger2016}. Interestingly, this quest has inspired the design of new materials and devices~\cite{Kim2007,Dogic2017,Mijalkov2016,Maggi2016,Aubret2018,Ross2019}.
 
Like the many artificial active systems recently proposed to tackle this question~\cite{Deseigne2010,Wang2013,Sanchez2012,Palacci2013,Bricard2013}, assemblies of motile bacteria turned out to be a rich and insightful experimental playground~\cite{Wu2000,Sokolov2010,DiLeonardo2010,Sokolov2012,Poon2013,Petroff2015,Lopez2015,Schwarz2012}. Among the rich topics that were investigated, the confinement of bacteria and of active particles has been the focus of many experimental~\cite{Wioland2013, Wioland2016, Creppy2016,Lushi2014} and theoretical studies~\cite{Theillard2017,Lushi2014}, showing that, under strong confinement, vortical collective motions may spontaneously appear. 

A class of bacteria ---called magnetotactic (MTB)--- can grow internally a microscopic magnet, hence providing an external handle to drive their swimming orientation~\cite{Blakemore1975,Uebe2016}.
As a source of nano-magnetic particles widely used in a medical context, MTB are micro-organisms of strong practical interest~\cite{Schuler1999}. For example, the magnetic alignment, combined with a micro-aerotactic swimming response, qualifies such micro-swimmers as a promising vector for targeted drug therapy~\cite{Houle2016}. Recently, it was proposed, on theoretical grounds, that a suspension of such magnetotactic bacteria could display original magneto-rheological properties~\cite{Vincenti2018, Alonso2018}, novel pattern formation~\cite{Guzman2016} and hydrodynamic instabilities \cite{Meng2018,Koessel2019}. In particular, the pearling hydrodynamic instability reported by Waisbord \textit{et al.}~\cite{Waisbord2016}, the velocity condensation \cite{Rupprecht2016} and the emergence of new phases induced by a magnetic field~\cite{Pierce2018} are striking examples of these.

Here, we study aqueous spherical droplets suspended in oil and containing a suspension of magnetotactic bacteria. We show how MTB self-assemble into a rotary motor under the application of a uniform and constant magnetic field, providing a mechanical torque to the fluid outside the droplets. 
In the self-assembly process, the magnetic field induces an accumulation of the swimming bacteria in diametrically opposed areas at the surface of the droplet. At high bacterial concentration, the flows resulting from the swimming activity of the bacteria and originating from these areas, interact to create a collective solid-like vortex flow in the central droplet core. Through Particle Image Velocimetry (PIV) analysis and particle tracking, we quantify the flows inside and outside the droplet and measure the net torque produced by this micromotor as a function of the magnetic field and the droplet radius. Finally, we provide an explanation on how an external torque can be generated despite the fact that the swimmers self-propel at almost zero Reynolds number.

\section*{Results.}

\subsection{Experimental setup.}
A water-in-oil emulsion is prepared by shaking a mixture of hexadecane oil with a suspension of magnetotactic bacteria (\textit{Magnetospirillum gryphiswaldense} MSR-1) (See Fig.~\ref{setup} (a) and Supplementary Movie 1). With our preparation protocol for bacteria~\cite{Schuler2014} (see methods and Supplementary Note 3 for details), MTB swim with a velocity $V_{0}$ ranging from \SI{20}{} to \SI{40}{\micro\meter.\second^{-1}} and exhibit a magnetic moment $m\sim$ \SI{e-16}{\joule.\tesla^{-1}}. The droplets encapsulate an almost even population of \textit{north-seeker} (NS) and \textit{south-seeker} (SS) bacteria, meaning that, under the application of a magnetic field, roughly half of the population will swim persistently towards the (magnetic) north and the other half towards the south. The emulsion is placed between two glass slides on the stage of an inverted microscope and at the center of a pair of Helmholtz coils, where a constant horizontal magnetic field is generated. The droplet radius, $R$, spans typically from \SI{20}{} to \SI{120}{\micro\meter}. Once the emulsion is formed, all the bacteria dwell in the aqueous phase. We call north pole (NP) the point on the droplet surface corresponding to the far-most position in the direction of the magnetic north, and south pole (SP) the diametrically opposed position (see Fig.~\ref{setup} (c)).

\subsection{Vortex flow inside the droplets.}
In  absence of magnetic field, regardless of the bacteria density, the swimming direction of MTB in the drops is random and unbiased. When the suspension is dense no collective motion is observed at the droplet larger scale; only fluctuating and intermittent vortices appear, with typical sizes much smaller than the droplet diameter. Note that the experiments of Wioland {\em et al.}\ \cite{Wioland2013} show the development of global vortices taking place at higher volume fraction.

\begin{figure}[h]
\centering
\includegraphics[width=14cm]{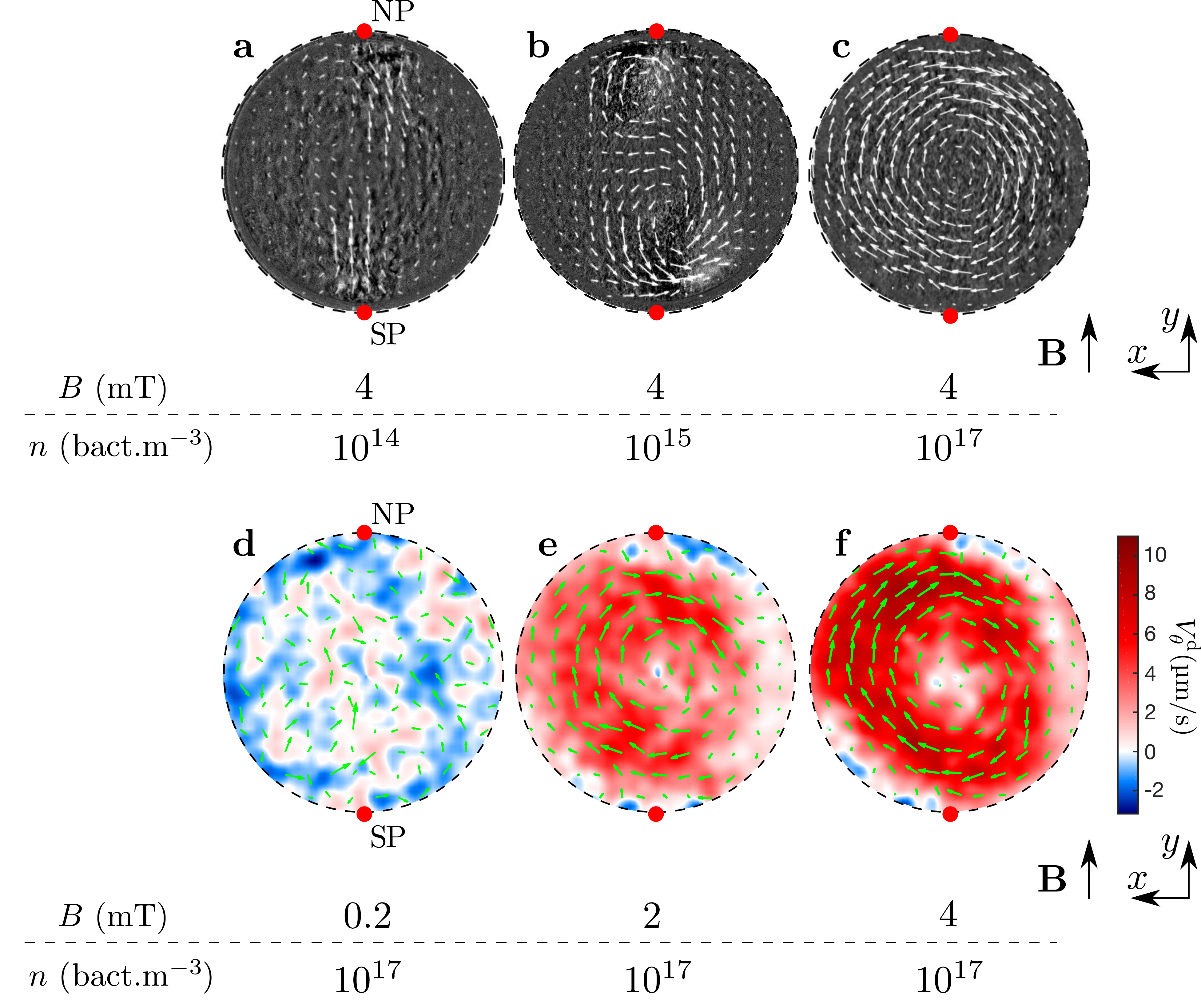}
\caption{\textbf{Influence of the cell density $n$ and the magnetic field $B$ on the emergence of collective vortical motion.} 
(a-b-c) $\times 40$ phase-contrast images of droplets superimposed with time average PIV velocity fields (green arrows). We show the influence of the cell density $n$ on the phenomenology, the magnetic field is fixed $B=$ \SI{4}{\milli\tesla}. (a) $n\sim \SI{e14}{bact.\meter^{-3}}$, $R=$ \SI{43}{\micro\meter}: Bacteria accumulate at the poles of the droplet. (b) $n\sim \SI{e15}{bact.\meter^{-3}}$, $R=$ \SI{89}{\micro\meter}: unstable recirculation flows appear at the poles of the droplet. (c) $n\sim \SI{e17}{bact.\meter^{-3}}$, $R=$ \SI{55}{\micro\meter}: the bacteria self-organize to form a stable vortex flow at the center of the droplet.
(d-e-f) Colored maps of the orthoradial projection of the instantaneous PIV velocity fields $V^{\rm d}_{\theta}$ (red-blue colormap, enhancing positive and negative values) superimposed with the instantaneous PIV velocity field (green arrows). The radius of the droplet is constant $R=$ \SI{83}{\micro\meter}. We show the influence of the magnetic field magnitude on the phenomenology, the cell density is fixed $n=\SI{e17}{bact.\meter^{-3}}$. (d) $B=$ \SI{0.2}{\milli\tesla}: no large scale collective motion is observed. (e) $B=$ \SI{2}{\milli\tesla}: vortex flow centered at the droplet center. (f) $B=$ \SI{4}{\milli\tesla}: the vortex flow is stronger than at $B=$ \SI{2}{\milli\tesla}. (e-f) Recirculation flows (negative values of $V^{\rm d}_{\theta}$) close to the poles are identified in blue.}
\label{phenomenology}
\end{figure}

When a magnetic field is set and in dilute conditions (for a cell density $n\sim\SI{e14}{bact.\meter^{-3}}$), NS (resp. SS) bacteria accumulate in the vicinity of the north (resp. south) poles of the droplet as a consequence of the bacteria swimming persistence described above (see Fig.~\ref{phenomenology} (a)). Because NS and SS bacteria are performing reversals, we observe some bacteria escaping from the accumulation regions (see Supplementary Movie 2). At an intermediate density (typically $n\sim\SI{e15}{bact.\meter^{-3}}$), the accumulation pattern becomes more unstable with episodic formation of jets propelling the fluid and the bacteria out of the polar positions, thus creating two local recirculation zones near each pole (see Supplementary Movie 3 and Fig.~\ref{phenomenology} (b)). For a dense suspension ($n\sim \SI{e17}{bact.\meter^{-3}}$), which is the case in the rest of the study, a steady and uniform collective rotational motion is observed, with an axis of rotation perpendicular to the magnetic field (see Supplementary Movie 4 and Fig.~\ref{phenomenology}~(c)) and oriented along the gravity direction. Although all symmetric planes containing the magnetic field direction could have been chosen by the bacteria, the vortex is actually rotating in the $x$-$y$ plane. This might be due to a sedimentation process (about $20 \%$ denser than the medium~\cite{Fan2015}) which yields a stable stratified suspension.
Visualization in the other horizontal planes shows a similar rotation field of equal direction as in the equatorial plane (see Supplementary Movies 5 and 6). The rotation direction chosen by the fluid is not completely random, with approximately $84\%$ of the drops rotating in a clockwise (CW) direction looked from the top. This preferential choice of spontaneous rotation is not completely elucidated yet but may be related to the helicity of MTB. 
An interesting property of this collective rotational motion is that, regardless of the choice of rotation direction at the magnetic field onset, the direction can be reversed by reversing the magnetic field (see Supplementary Movie 7). 
Some experiments were performed with a larger fraction of NS bacteria, selecting them using a macroscopic magnet. At the scale of the droplets, some SS individuals are recovered, but in significantly less quantity  than the NS. Even with such an unbalanced ratio of NS/SS, we found that the collective rotation is still preferentially CW.

From now on, we focus on the characteristics of the vortex flow at a fixed density $n\sim \SI{e17}{bact.\meter^{-3}}$. For a magnetic field larger than a threshold value (typically $0.4 \pm 0.1$ \SI{}{\milli\tesla}), one observes the emergence of the large scale vortical flow previously mentioned. At the experimental density, individual bacteria are not observable, however, through PIV analysis of phase-contrast microscopy images, we obtain the temporally and spatially resolved velocity fields due to the bacterial motion inside the drops $\mathbf V^{\rm d}(x,y)$ (Fig.~\ref{phenomenology}~(d-e-f)). The flow geometry is reminiscent of the experiments by Sokolov \textit{et al.}\ \cite{Sokolov2016}, but here the bacteria are aligned along the magnetic field. The flow field shows a central vortical structure and presents two maximal streams located near the poles reminiscent of the two jets visualized at lower bacterial concentration. The strength of the vortical flow field increases with the intensity of the magnetic field (Fig.~\ref{phenomenology}~(e-f)). 
Computation of the angular average of the orthoradial velocity, $\overline{V^{\rm d}_{\theta}}(r)=\frac{1}{2\pi}\int_0^{2\pi} \! V^{\rm d}_{\theta}(r,\theta) \, \mathrm{d}\theta $, brings evidence for an effective  solid-core rotating motion, characterized by an angular velocity $\Omega^{\mathrm{d}}$ ($\overline{V^{\rm d}_{\theta}}(r)= \Omega^{\mathrm{d}} r$, see Fig.~\ref{motor_characteristics}~(a)). The solid-core spans one-half of the droplet radius for all the radii investigated. At increasing magnetic field intensities, the magnitude of $\Omega^{\mathrm{d}}$ increases to saturate at larger magnetic fields. Beyond the droplet core, the suspension is sheared and the velocity decreases down to a non-zero value at the droplet interface. 
PIV analysis also shows local recirculating regions in a direction opposite to the core rotation (see blue regions enhanced in velocity maps on Fig.~\ref{phenomenology} (e-f)), shifted with respect to the poles in the direction of the rotating motion.

\subsection{Flow in the oily phase.}
By tracking \SI{1}{\micro\meter}-diameter melamine resin beads in the surrounding hexadecane oil, we observe a net circular flow outside the droplets (see Supplementary Movie 4 and Fig.~\ref{motor_characteristics}~(b)), indicating the outcome of a net torque on the fluid outside the droplet. Indeed, although zero-torque sources can generate finite circulations, it can be proven that the circulation must change sign when measured at different planes (see Supplementary Note 4). As shown in Supplementary Movies 4, 5, and 6, the circulation sign is the same for all planes, which can only be produced by torque sources. Also, zero-torque sources would produce mainly radially oriented flows (eq.~4 of the Supplementary Note 4), contrary to the orthoradial flows we measure. 
Hence, the magnetotactic bacteria self-assemble inside the droplet to form a rotary motor. The angular average of the orthoradial velocity of the tracers $\overline{V_\theta^{\rm oil}}(r)$ is determined for different outer radii $r$ and for various magnetic field $B$ and droplet radii $R$ (see methods for details). In all cases, we measured a net fluid rotation in the same direction as the central core rotation. However, we observe that local recirculation patterns, opposed to the net fluid rotation, appear close to the poles, mirroring the previously mentioned counter-flow inside the droplet (see Supplementary Movie 4). 

\subsection{Torque measurements.}
\begin{figure}[h]
\centering
\includegraphics[width=13cm]{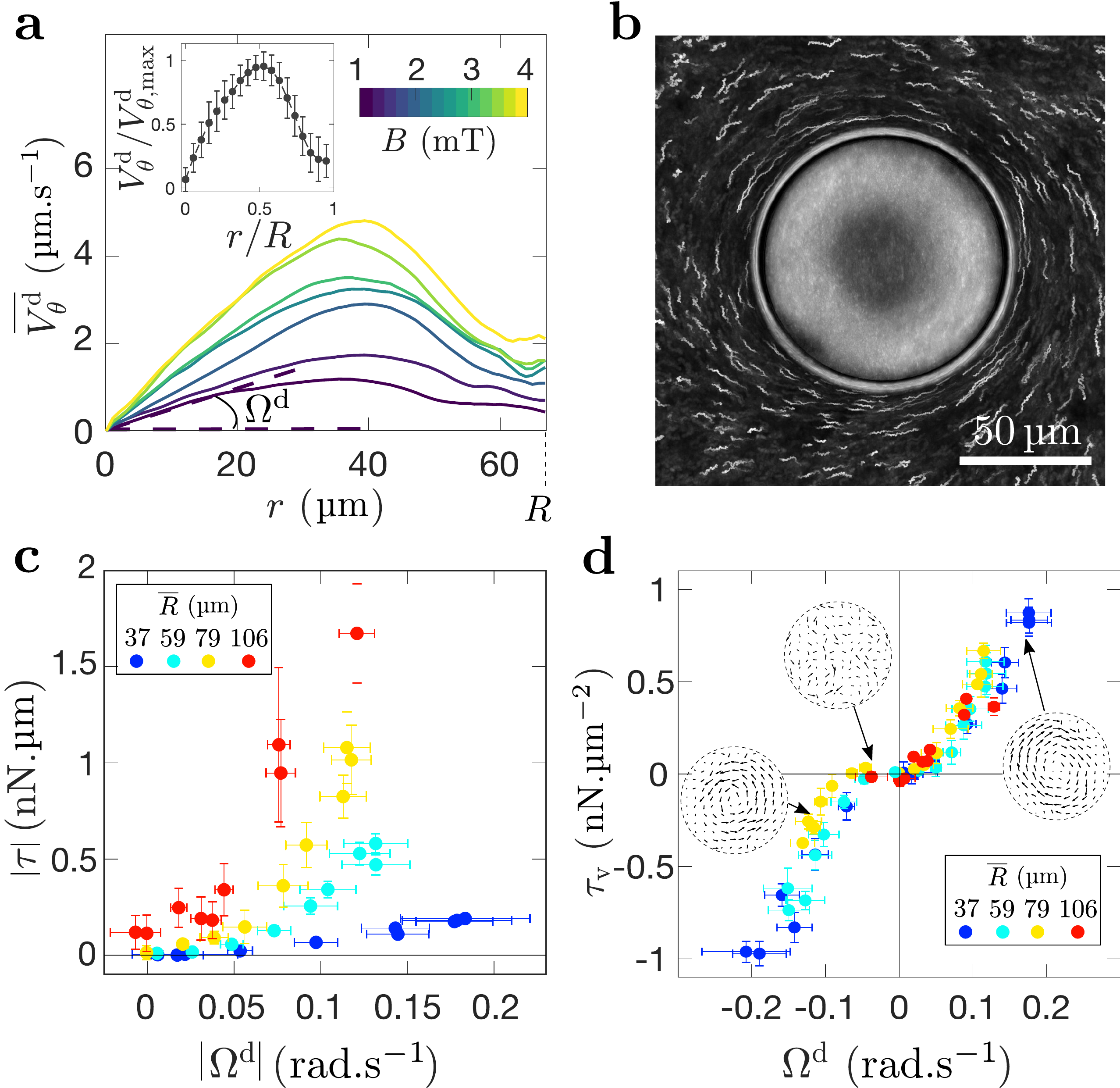}
\caption{\textbf{Mechanical characterization of the rotary motor.} (a) Mean orthoradial velocity profile $\overline{V^{\mathrm{d}}_{\theta}}(r)$ for one droplet of radius $R=\SI{67}{\micro\meter}$ and for different magnetic field magnitudes $B$ (colors, from bottom to top $B=$ 1, 1.4, 2, 2.4, 3, 3.4, 4 \SI{}{\milli\tesla} ). Close to the droplet core ($r=0$), the suspension rotates like a solid with a characteristic rotational velocity $\Omega^{\mathrm{d}}$ which increases with $B$. The errorbars are the standard errors. (b) Superposition of phase-contrast images (350 images corresponding to a \SI{14}{s} movie) showing the circular rotation of the outer tracers for an inner rotational velocity $\Omega^{\rm d}= \SI{0.13}{\radian.\second^{-1}}$ measured at $B=\SI{4}{\milli\tesla}$. (c) The torque $\tau$, acting on the oil and produced by the droplets, is extracted from the tracers orthoradial velocities (see (b)). We measure $\tau(B,R)$ for different droplets radii $R$ and magnetic field $B$ with respect to the core rotation velocity $\Omega^{\rm d}(B,R)$ (average data for 10 droplets of similar radii, $\overline{R}$ is indicated by colors and is given with a $\pm15$ \SI{}{\micro\meter} standard deviation). The errorbars are $\sigma/\sqrt{N}$ where $\sigma$ is the standard deviation and $N=10$ (number of droplets used for the average). (d) Torque by unit volume $\tau_{v}=\tau/(\frac{4}{3}\pi R^{3})$ as a function of $\Omega^{\rm d}$ for the same data set. The average data for different $\overline{R}$ collapse on the operating curve of the rotary motor. The velocity maps are the ones of the droplet displayed on Fig.~\ref{phenomenology}~(d-e-f) and placed at the corresponding operating points. The errorbars are the standard errors.}
\label{motor_characteristics}
\end{figure}

In the following, we measure the energy production associated with the rotary motor (i.e., the effective torque acting on the oil) and identify the mechanism of the torque generation inside the droplets. The effective torque exerted by the rotary motor is extracted by fitting the radial dependence of the mean orthoradial velocity in the oil phase with a simple hydrodynamic model (see methods and Supplementary Note 1), consisting in a sphere driven in rotation by a torque $\tau$. However, the drop being sedimented at the bottom of the chamber, a hydrodynamic image of the rotating droplet is added to account for the no-slip boundary condition of the flow field at the solid interface. The dependence of the orthoradial projection of the external flow, $\overline{V_\theta^{\rm oil}}=\frac{1}{2\pi}\int_{0}^{2\pi} \! V^{\rm oil}_{\theta}(r,\theta) \, \mathrm{d}\theta $, with the distance $r$ to the center of the droplet reads (for $r>R$):
\begin{equation}
\overline{V_\theta^{\rm oil}}(r) = \frac \tau{8\pi \eta_{\mathrm{oil}}R^2} \left[\frac{R^2}{r^{2}} - \frac{r/R}{(r^2/R^2 + 4)^{3/2}} \right],
\label{TorqueEq}
\end{equation}
where $\eta_{\rm oil}=$ \SI{3e-3}{ {Pa.s}} is the dynamic viscosity of hexadecane at 25$^{\circ}$C. The dependency with $\frac{R^2}{r^2}$ results from the rotation of a sphere of radius $R$ in the bulk and $- \frac{r/R}{(r^2/R^2 + 4)^{3/2}}$ is a correction term accounting for the presence of the bottom plate of the pool.
From the experimental measurement of $\overline{V_\theta^{\rm oil}}(r)$, we are then able to estimate $\tau$ for various droplets radii at different magnetic field intensities. Fig.~\ref{motor_characteristics}~(c) shows the dependency of $\tau$ with respect to the core solid rotation of the MTB suspension $\Omega^{\rm d}$ for different mean radii (each data point corresponds to an average over several droplets of similar radii): we observe that $\tau$ increases with $\Omega^{\rm d}$ and with the droplet radius. Similarly, we plot on Fig.~\ref{motor_characteristics}~(d) the torque by unit volume $\tau_{v}=\tau/(\frac{4}{3}\pi R^{3})$ which appears to collapse all the data onto a unique curve. This curve corresponds to the operating curve of the droplet motor, analogous to the ones of macroscopic synchronous motors, pointing out a direct link between the core rotation and the flow generation outside the droplet. This collapse means that $\tau_{v}$ and $\Omega^{\rm d}$ have a similar dependency with respect to the parameters of our experiments, meaning the droplets radii and the magnetic field intensities. 
Within experimental uncertainties, the inversion symmetry of this curve is clearly visible with possibly a small bias towards CCW rotation. This implies that the symmetry breaking mechanism acts mainly at the selection of the direction of rotation, but only weakly on the operation.
The non-linear shape of this operating curve shows that the motor is less efficient at low $\Omega^{\rm d}$ (typically for $\Omega^{\rm d}<$ \SI{0.05}{\radian.\second^{-1}}) than at high $\Omega^{\rm d}$. 
At low $\Omega^d$, corresponding to a low magnetic field, the motor is less efficient because the inner vortex structure is not well established. On top of this, Brownian motion on the outer tracers makes the torque measurements quite noisy.

\subsection{Mechanism of torque generation.}
  A global circulation, resulting from a net torque acting on the oil phase and produced by torque-free and force-free swimmers, can only be sourced in the misalignment dynamics of the magnetic moments of the MTB with the external magnetic field.
 In a quiescent fluid, bacteria align with the field in a time $t_B = \xi_r/(mB)$, where $\xi_r=\pi\eta_{w} \ell^3/[3 \ln(2\ell/a)]$ is the rotational friction coefficient, $l=3-4$ \SI{}{\micro\meter} is the body length of one bacterium, $a=\SI{1}{\micro\meter}$ its width and $\eta_{w}=$ \SI{1e-3}{ {Pa.s}} is the dynamic viscosity. With these typical values, we find that $t_B\approx\SI{0.1}{\second}$ for $B=\SI{2}{\milli\tesla}$. 
When bacteria swim close to the droplet interface, they are forced to turn in order to align along the interface before swimming parallel to the boundary (this alignment has been clearly observed for dilute suspensions, see Supplementary Note 2 and Supplementary Movie 8). The time needed for one bacterium to orient along the boundary is $t_T=\ell/V_0\approx$~\SI{0.1}{\second}, which is of the same order of magnitude as $t_B$. Hence, the droplet boundary makes the bacteria at the interface misaligned with the magnetic field, which leads to the generation of torque in the droplet. This misalignment depends on the position of the bacteria in the droplet and is expected to be the strongest at the north and south poles of the droplet, where the interface is perpendicular to the magnetic field direction.
To precise the kinematics involved inside the droplet, we can take the example of a droplet rotating CW without loss of generality. In this case, due to the advection by the central vortex core and to the swimming propulsion along the magnetic field direction, NS (resp. SS) bacteria reach the droplet boundary and align with it at the right (resp. left) of the NP (resp. SP) in the $x$-$y$ plane. The magnetic torque on those misaligned bacteria (both at the NP and SP) points in the CW direction, thus reinforcing the circulation. 
As NS (resp. SS) bacteria swim along the droplet boundary, their misalignment with $\mathbf{B}$ increases while getting close to the NP (resp. SP).
Then, trespassing the NP (resp. SP), the situation  becomes unstable  because the magnetic torque is too large to be compensated by the boundary alignment and also, because they will meet a flow of bacteria transported by the global rotation. 
Then, these  bacteria changing orientation will leave the counter rotating droplet boundary to be advected by the vortical flow.
This orientation flip will, most probably, cause a strong release of magnetic torque in the fluid.

\begin{figure}[h]
\centering
\includegraphics[width=13cm]{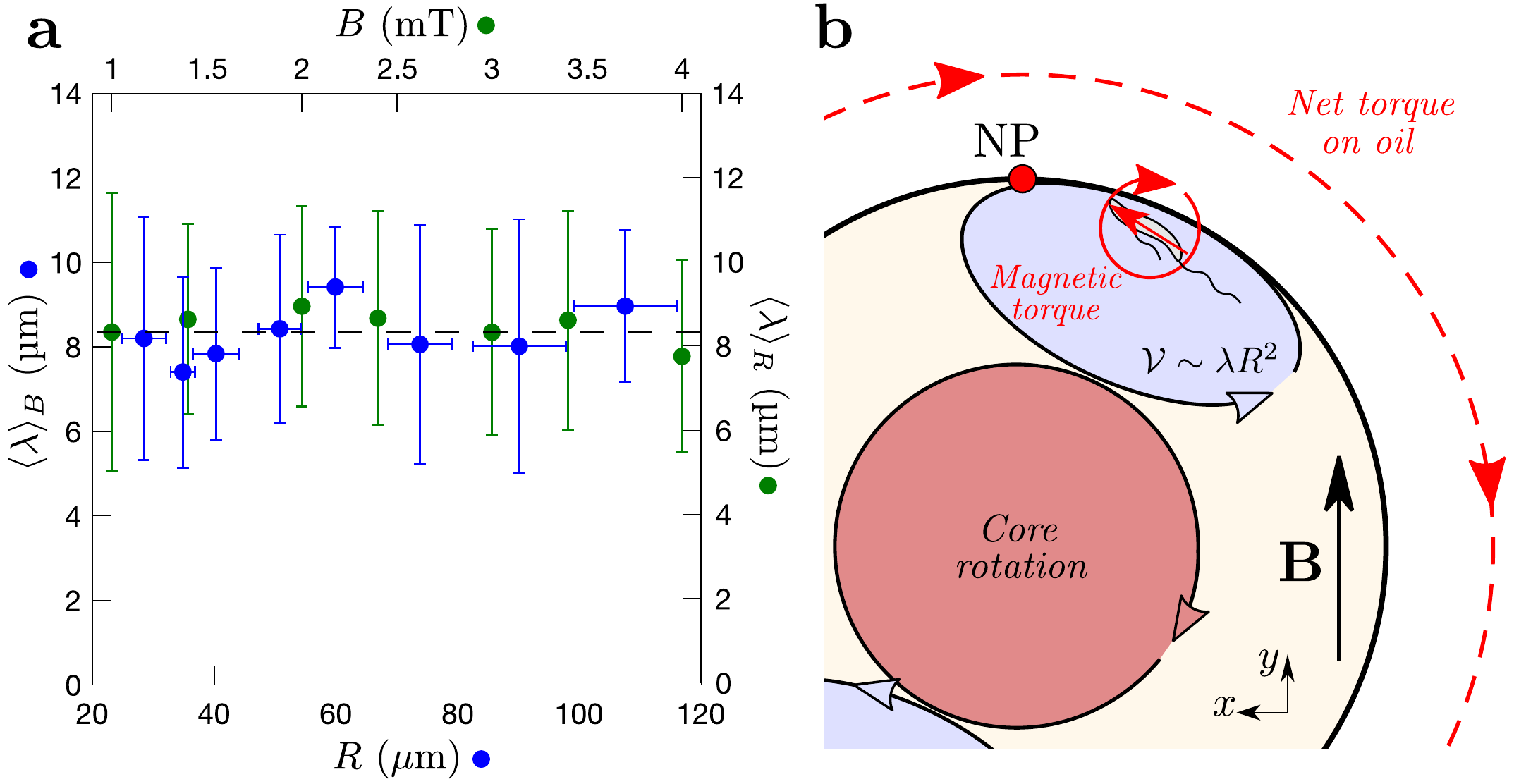}
\caption{\textbf{Test of the scaling relation: $\tau=nmB\lambda(R,B)R^{2}$}, where $\tau$ is the generated torque, $n\sim \SI{e17}{bact.\meter^{-3}}$ is the bacteria density, $m\sim \SI{e-16}{\joule.\tesla^{-1}}$ is the magnetic moment of a single bacterium, $B$ is the magnetic field intensity and $\lambda(R,B)$ is a typical length inherent to the torque generation. (a) $\langle\lambda\rangle_{B}$ (resp. $\langle\lambda\rangle_{R}$) is the average value of $\lambda$ with respect to $B$ (resp. $R$). We only included the data corresponding to  $\tau_{v}> \SI{0.1}{\nano\newton.\micro\meter^{-2}}$, for which the torque is strong enough to be measured out from experimental noise. The error bars are the standard deviations of the average data. This graph shows that $\lambda=8\pm 2$ \SI{}{\micro\meter} is an intrinsic length of the system which does not depend on $R$ nor $B$. The errorbars are the standard errors. (b) Qualitative interpretation of the rotary motor self-organization. The volume of the recirculating bacteria contributing to the torque is dimensionally $\mathcal{V}\sim\lambda R^{2}$. The picture is similar close to the south pole of the droplet.}
\label{characteristic_length}
\end{figure}

This picture is consistent with the previously mentioned counter-rotating flows inside and outside the droplet.
 From this time-scale analysis, we can therefore infer that the net torque is produced by the bacteria misaligned with the magnetic field at the droplet boundary, which points on a surface effect. More precisely, we are able to state that:
 \begin{equation}\label{motor_torque}
 \tau=n\mathcal{V} mB,
 \end{equation}
where $\mathcal{V}$ is the volume of bacteria contributing effectively to the torque. To account for surface effects, we expect $\mathcal{V}\sim \lambda R^{2}$, where $\lambda$ is a typical length independent of $R$. Indeed, by computing $\lambda=\tau/(nR^{2} mB)$ (see Fig.~\ref{characteristic_length}~(a)), we bring evidence of a characteristic length $\lambda=8\pm2$ \si{\micro\meter} that does not depend neither on the magnetic field intensity nor on the droplet radius, collapsing all the data we collected. $\lambda$ is of the order of a bacterium size and can then be related to a microscopic scale, consistent with no dependence on neither $R$ nor $B$. Then, the picture emerging from our scaling analysis is that of one core rotation and two counter-rotating regions made of self-assembling bacteria at the poles and yielding a net torque to the oil (see Fig.~\ref{characteristic_length}~(b)). 
The active torque generated in the droplet $\tau$ would then correspond to an engine mechanical power  estimated through the flow generated in the oil phase as: $\mathcal{P} \approx \tau^2/(\eta_{oil} R^3)$. For a typical torque of $\tau = \SI{1}{\nano\newton\micro\meter}$,  oil viscosity $\eta_{oil} \approx \SI{3e-3}{\pascal.\second}$ and for a radius $R=\SI{80}{\micro\meter}$, the estimation yields  $\mathcal{P} \approx \SI{e-16}{\watt}$.

\subsection{Vortex reversal.}
\begin{figure}[h]
\centering
\includegraphics[width=16cm]{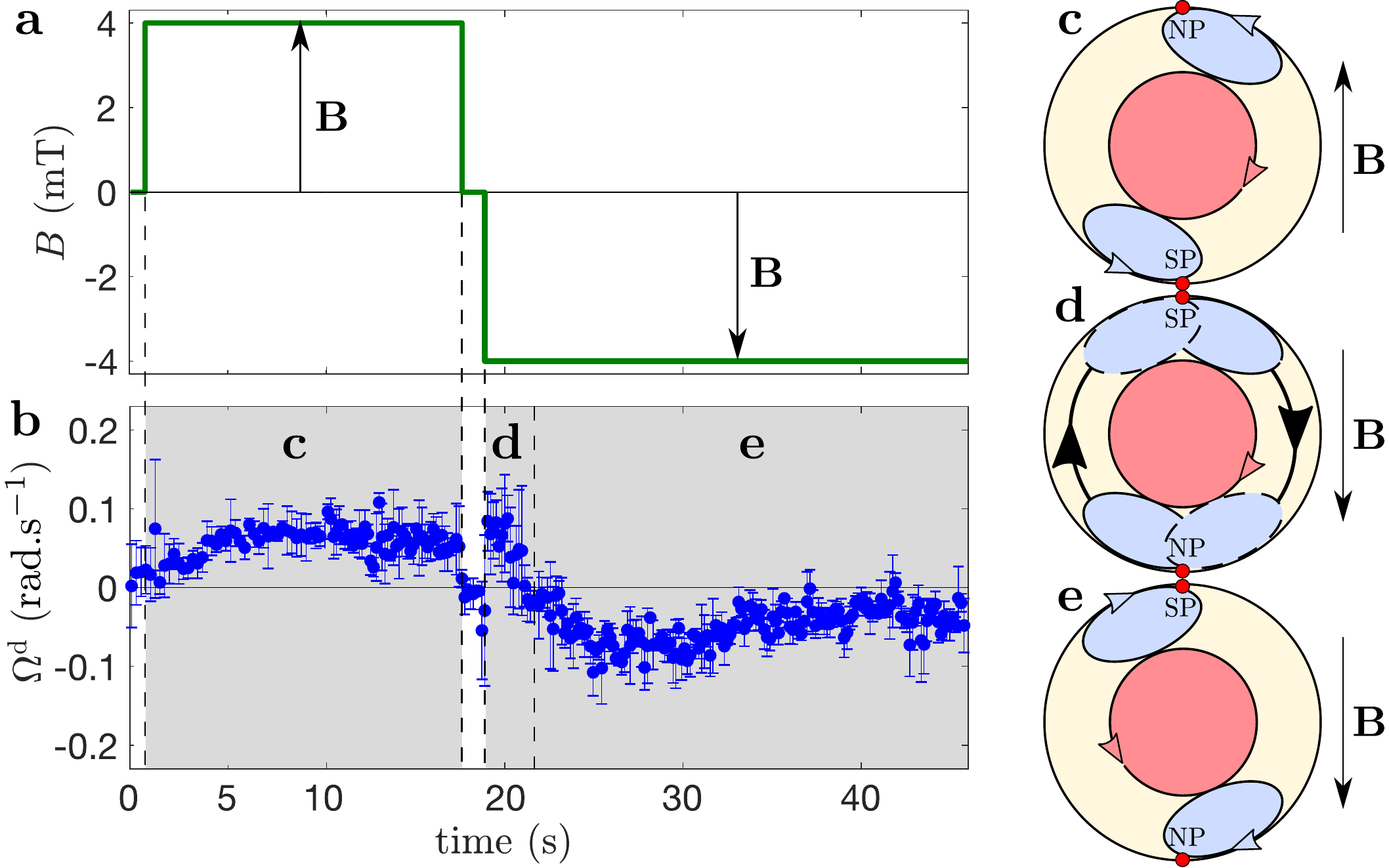}
\caption{\textbf{Vortex emergence and rotation reversal by magnetic field inversion.} (a) magnetic field amplitude $B$ as a function of time. (b) response in the rotational velocity of the droplet core $\Omega^{\rm d}$ under the applied magnetic field. The errorbars are the standard errors. (c), (d) and (e) are the stages corresponding to the particular self-assemblies representations on the sketches on the right of the figure. The coral region corresponds to the central core rotation while the blue zones represent the recirculating bacteria. The bacterial suspension is dense ($n\sim \SI{e17}{bact.\meter^{-3}}$). For each time step $t$, $\Omega^{\rm d}(t)$ is computed from instantaneous PIV maps as in Fig.~\ref{motor_characteristics}~(a). From the moment when the magnetic field is set on, $\Omega^{\rm d}$ reaches a stationary value within a few seconds ($\sim\SI{4}{\second}$). When reversing quickly the magnetic field direction while the suspension rotates CW, the suspension continues rotating CW at the short times after reversal before reversing completely its rotational direction to CCW. Then, $\Omega^{\rm d}$ reaches a stable negative value $\sim$\SI{10}{\second} after the magnetic field switch.}
\label{reversal}
\end{figure}

From our understanding of the bacteria self-assembly inside the droplet, we can provide an explanation of the rotation reversal actuated by the magnetic field switch mentioned above. When the magnetic field is rapidly reversed, the MTB that are accumulated at the right of the NP and at the left of the SP can suddenly rotate to align along the new magnetic field, without any restriction from the droplet boundary. After this, they cross the droplet roughly along the $y$-direction, producing a transient convective circulation with positive values of $\Omega^{\rm d}$ (corresponding to CW direction) measured just after the magnetic field reversal (see Fig.~\ref{reversal} and Supplementary Movie 9). Then, they accumulate near the new NP and SP, which are located on opposite positions to the original ones. In this process, the MTB that were accumulated CW of their respective poles are now CCW of the new poles and, hence, generate an opposite torque provoking the inversion of the motor (see the full quantitative report of this dynamics on Fig.~\ref{reversal}). For this mechanism to be efficient and overcome the natural tendency to generate a CW motor, the time duration between the application of the two opposite magnetic fields must be shorter than the thermal reorientation time (of the order of the inverse rotational diffusion constant, $D_r=1/80$ \si{\second^{-1}}), as it is indeed the case on Fig.~\ref{reversal}. Otherwise, the MTB have time to reorient isotropically and leave the accumulation zones.

\section*{Discussion.}
Recently, considerable efforts have been undertaken to harness the microscopic activity of living or synthetic agents like bacteria~\cite{Vizsnyiczai2017,Schwarz2012}, eukaryotic cells~\cite{Williams2014}, Janus colloids~\cite{Aubret2018} or micro-robots~\cite{Deblais2018,Scholz2018}, in order to extract macroscopic work from microscopic mechanical structures. Here, we show a remarkable example of living biological entities self-assembling into a rotary motor actuated by a controlled, external aligning field. 
In this process of self-organization, confinement plays an essential role. It would probably be fruitful to exploit it by an adapted design of a micro-fluidic environment. 
To obtain such a complex organization, a first key-point is the ability for magnetotactic bacteria to accumulate in some areas (here being close to the droplet poles) under a uniform magnetic field. This accumulation of active swimmers, obviously limited by crowding, results in an instability triggering a coordinated motion at the scale of the droplet.
The second crucial point relies on the original properties of this active magnetic fluid. Any swimming kinematics inducing a change in the magnetic moment orientation of the bacteria with respect to the applied magnetic field, is bound to produce torque on the surrounding fluid. Interestingly, this situation is reversed compared with the standard synchronous motor, exemplified by the magnetic stirrer of chemical labs, which follows the rotation of a rotating magnetic field. In our case, a swimming bacterium carrying a magnetic moment is rotated and creates a torque due to the confinement in a droplet, while the magnetic field direction remains fixed.
Even though the physical origin of the self-organization process is not completely elucidated yet, we expect a similar behaviour for other types of autonomous swimmers, confined and orientable by any external field (electric field, light,...), providing new routes of theoretical and experimental investigations.

\section*{Methods.}
\noindent{\bf Bacteria growth protocol.} 
We used magnetotactic bacteria (MTB) from the MSR-1 \textit{Magnetospirillum gryphiswaldense} strain. MTB are grown in a {Flask Standard Medium} (FSM) in the absence of external magnetic field, though the Earth magnetic field is still present. This medium was beforehand bubbled with a gas containing 2\% O$_{2}$ and 98\% N$_{2}$ and sealed inside Hungate tubes of 12 mL. We use inoculation volume of 300 $\mu$L to start bacteria growing in a tube. In such conditions, we got roughly 50\% of NS and 50\% of SS MTB in the suspension, consistent with a standard growth protocol of MSR-1~\cite{Schuler2014}. Bacteria used for the experiments shown here are harvested at the end of the growing sigmoid in order to work with the most motile swimmers (bacteria concentration corresponding to an optical density OD$=0.12\pm 0.02$ measured at a wavelength of \SI{600}{\nano\meter}). 

\begin{figure}[h]
\centering
\includegraphics[width=13cm]{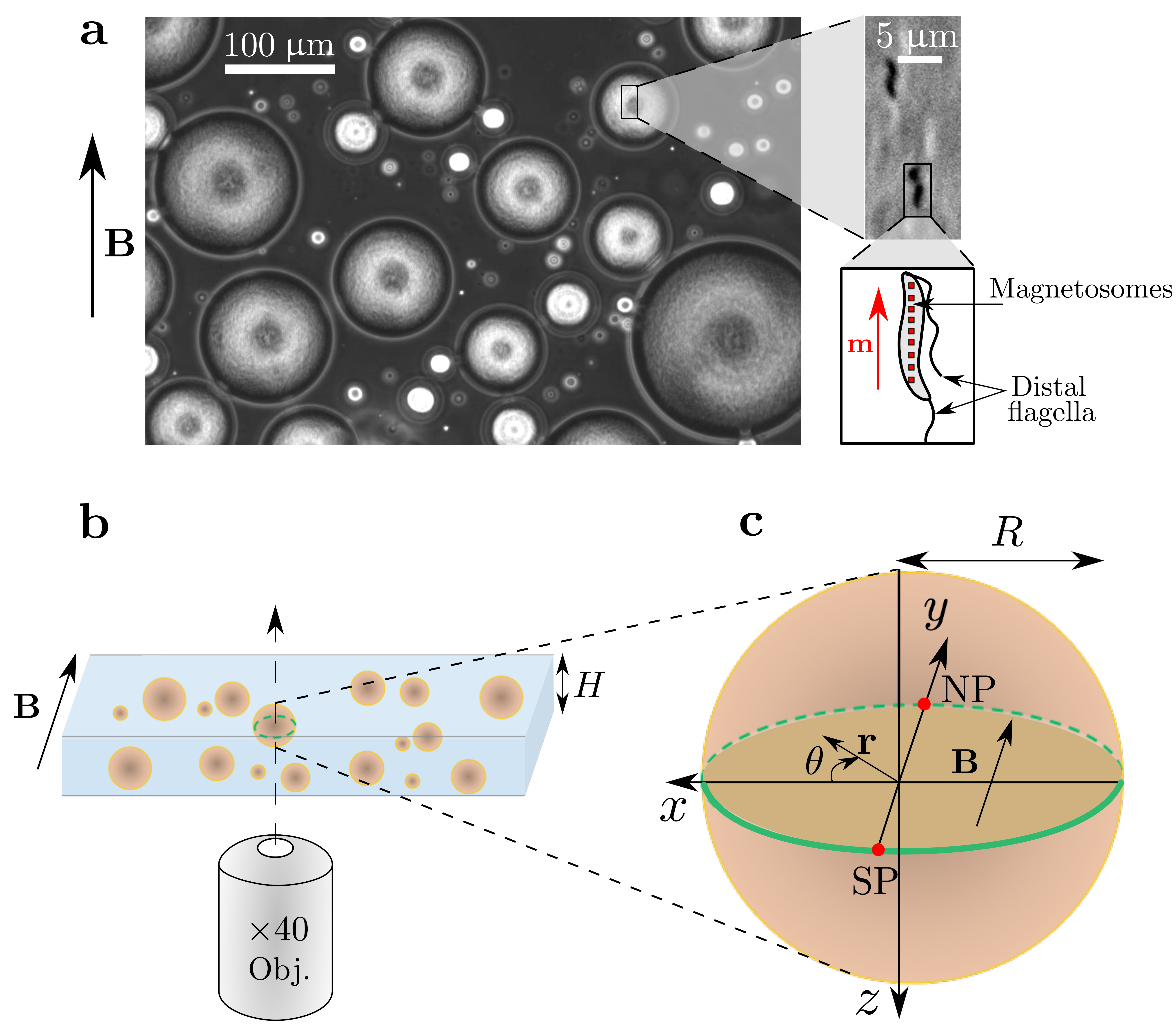}
\caption{\textbf{Water-in-oil emulsion of magnetotactic bacteria.} (a) $\times 10$ phase-contrast image of an emulsion of magnetotactic bacteria (bacteria remain inside droplets) in hexadecane oil. A magnetic field  of \SI{4}{\milli\tesla} is applied as indicated by the arrow (see the corresponding Supplementary Movie 1). A broad distribution of droplets radii is obtained, spanning typically from 20 to 120 \SI{}{\micro\meter}. \textit{Zoom in:} $\times 40$ phase-contrast image of two magnetotactic bacteria \textit{Magnetospirillum gryphiswaldense} MSR-1 (darkest zones) swimming along the magnetic field direction. \textit{Zoom in:} Sketch of a magnetotactic bacterium carrying magnetosomes (red squares) and two distal flagella. The magnetosomes are aligned along the body, generating a magnetic moment $m$. (b) Setup principle: a droplet, lying on the bottom plate of a pool of height $H=$ \SI{270}{\micro\meter} and placed on the stage of an inverted microscope, is observed at its equatorial plane with a $\times 40$ objective. A uniform magnetic field is applied in the observation plane, parallel to the bottom and top plates of the pool. (c) Definitions of the north pole (NP), the south pole (SP) of the droplet and the spatial coordinates. $R$ is the droplet radius.}
\label{setup}
\end{figure}

\noindent{\bf Emulsion preparation and setup control.} 
The MTB are initially in a flask standard medium and eventually concentrated by centrifugation. Then, an emulsion is prepared by agitation in the presence of hexadecane oil (\textit{ReagentPlus}, SigmaAldrich) containing Span80 (2\%-weight concentration) as surfactant to stabilize the emulsion. We prepare samples at estimated bacterial density from $10^{14}$ to $\SI{e17}{bact.\meter^{-3}}$ (volume fractions of $0.01\%$ and $10\%$, using a bacterial body volume of $\mathcal V_b = \SI{3}{\micro\meter^{3}}$). This bacterial number density is estimated for a given OD (Optical Density), low enough to count the bacteria using phase-contrast images, hence giving a conversion between OD and bacteria concentration. A volume of \SI{65}{\micro\liter} of the emulsion is then deposited in a chamber composed of a double-sided tape adhered to a microscope slide and sealed with a glass cover slip on top (see Supplementary Note 3). This system creates a closed pool of area \SI{1.5}{\centi\meter} $\times$ \SI{1.6}{\centi\meter} and height $H =$ \SI{270}{\micro\meter}. The emulsion is visualized inside the pool using an inverted microscope adapted to receive Helmholtz coils, which produce a uniform horizontal magnetic field $\mathbf{B}$ (see Fig.~\ref{setup} (a)). That is, the visualization plane ($x$, $y$) is parallel to $\mathbf{B}$.  The intensity of the magnetic field $B = |\mathbf B|$ is controlled electronically, from 0 to \SI{4}{\milli\tesla} with a precision of \SI{0.1}{\milli\tesla}. Droplets of diameters smaller than \SI{270}{\micro\meter} sediment at the bottom surface of the chamber due to the low density of hexadecane oil. For a given emulsion preparation, several droplets are visualized in their equatorial plane with respect to the vertical direction (See Fig.~\ref{setup} (b)). For this report, we use mostly a $\times 40$ phase-contrast objective (Zeiss A-Plan Ph2 Var2, mounted on a Zeiss AXIO Observer microscope) which allows full visualization of the droplets. For all the experiments, we only observe droplets sufficiently separated from each other (typically distant of, at least, one droplet diameter) to avoid any coupling effects between droplets. Experiments are always performed within 30 minutes after centrifugation for the largest bacteria concentration (longer observations have shown a decrease of bacteria motility after this time). For flow visualization outside the droplets, we used \SI{1.1}{\micro\meter} melamine resin beads suspended in hexadecane oil.

\noindent{\bf Data acquisition and analysis.}
Phase-contrast images are recorded using a Hamamatsu ORCA Flash4 camera equipped with a CCD sensor of $2048 \times 2048$ pixels. For movies, a frame rate of \SI{25}{\hertz} is chosen to capture the full dynamics inside and outside the droplets.  To prepare PIV analysis, we post-process raw images from experiments by subtracting the average-intensity image of a stack to all the images of the stack. This allows us to get rid of the intensity gradients (which could lead to discrepancies in the flow measurements) inherent to both phase-contrast microscopy and the spherical shape of the droplets. It also provides better accuracy on the velocity map close to the droplet interface. We choose interrogation window size to be equal to $32\times 32$ pixels (corresponding to $5\times 5$ \SI{}{\micro\meter}) and an overlap between windows equal to the half of the size of a window. A standard FFT cross-correlation algorithm is used to compute the PIV velocity field using the Matlab PIVlab facilities. We compare successive images separated by 1/25 s. In order to compute the average ortho-radial velocity profile $\overline{V^{\rm d}_{\theta}}(r)$, we average the velocity field on a movie of typically 350 images (14 s).
For tracking, we used the TrackMate plugin of Fiji (extension of ImageJ). To smooth thermal noise between two successive tracking points in time, we average the velocity of each tracked particle on two successive images. To get the orthoradial velocity field, we compute the circulation $\mathcal{C}(r,t)=\int_0^{2\pi} \! rV^{\rm oil}_{\theta}(r,\theta) \, \mathrm{d}\theta $ of the experimental velocity field on circles of various radii $r$ centered on the droplet center at time $t$. Then, this circulation is averaged in time on the duration of the movie (typically 14 s) to get the mean circulation $\overline{\mathcal{C}}(r)=\langle C(r,t)\rangle_{t}$. Then, we obtain the average orthoradial velocity field $\overline{V^{\rm oil}_{\theta}}(r)=\overline{C}(r)/2\pi r$, which is the net velocity of the outer flow. The advantages of this method are both to increase accuracy by smoothing Brownian motion of the tracers and to give a reliable estimate of the net torque applied by the droplet on the oil (counter-flows, opposite to the main recirculation flow, are taken into account in $\overline{V^{\rm oil}_{\theta}}(r)$).

\newpage

\subsection*{Acknowledgements}
The authors thank Xavier Benoit-Gonin for his creative technical help on the set-up and Thierry Darnige for his determinant assistance on the numerical  interfaces. E.C. is grateful to Prof. Dirk Sch\"uler for enlightening scientific discussions on MTB and on providing the MSR-1 strain. 
We acknowledge the support of the ANR-2015 BacFlow under Grant No.\ ANR-15-CE30-0013, Franco-Chilean EcosSud Collaborative Program C16E03,  Fondecyt Grants No.\ 1180791 and 1170411, and Millenium Nucleus Physics of Active Matter of the Millenium Scientific Initiative of the Ministry of Economy, Development and Tourism (Chile).

\subsection*{Author contributions} B.V. and G.R. did the experiments. B.V., G.R., M.L.C., C.D., R.S., and E.C.  participated to the scientific discussion and the writing of the article.

\subsection*{Competing interests} The authors declare no competing financial or non-financial interests.

\subsection*{Correspondence} Correspondence and requests for materials should be addressed to E.C. \\ (email: eric.clement@upmc.fr).

\subsection*{Data availability} The authors declare that the data supporting the findings of this study are available within the paper and its supplementary information files.

\cleardoublepage

\ 
\newpage

\begin{center}
\huge{Supplementary Information}
\end{center}

\subsection{Supplementary Note 1. Hydrodynamic model of a sphere rotating in a fluid close to a surface}
To estimate the torque exerted by a droplet of radius $R$ on the oil, we use an hydrodynamic model of a sphere of radius $R$ rotating in a fluid of viscosity $\eta=$ \SI{3e-3}{{Pa.s}} (value for hexadecane oil at 25$^{\circ}$C). The droplets being sat on the bottom plate of the pool, we have to take into account, in our model, the hydrodynamic image of the rotating sphere with respect to the bottom plate (see Supplementary Fig.~\ref{hydrodynamic_model}~(a-b)). The flow created by a sphere of radius $R$, rotating thanks to a torque $\tau$, reads, in the bulk :
\begin{equation}
V^{\rm oil,1}_{\theta}(r,\phi)=\tau\sin\phi/(8\pi\eta r^2),
\label{flow_bulk}
\end{equation}
where $\phi$ is the azimuthal angle, $\phi=\pi/2$ at the equator of the droplet.

The correction due to the bottom wall corresponds to the superimposition of the flow in Eq.~\ref{flow_bulk} and the flow created by the mirror image of the sphere with respect to the wall, which counter-rotates with respect to the real droplet rotation. The flow created by the mirror droplet at the equatorial plane of the real droplet is :
\begin{equation}
V^{\rm oil,2}_{\theta}(r',\phi')=-\tau\sin\phi'/(8\pi\eta (r')^2),
\end{equation}
with $r'=\sqrt{r^{2}+4R^{2}}$ and $\phi'= \pi - \arcsin\left( \frac{r}{r'} \right)$. Then, $\sin\phi'=r/\left(\sqrt{r^{2}+4R^{2}}\right)$ and the total flow at the equatorial plane of the droplet reads :
\begin{equation}
V^{\rm oil}_{\theta}(r)=V^{\rm oil,1}_{\theta}(r,\phi)+V^{\rm oil,2}_{\theta}(r',\phi'),
\end{equation}
\begin{equation}
\Rightarrow V^{\rm oil}_{\theta}(r)=\frac{\tau}{8\pi\eta R^{2}}\left( \left(\frac{R}{r}\right)^{2} - \underbrace{\frac{r/R}{((r/R)^{2}+4)^{3/2}}}_{\rm corrective \ term \ due \ to \ the \ bottom \ wall} \right),
\end{equation}
where the corrective term due to the bottom wall is emphasized.

The experimental flow field is very well captured by this hydrodynamic model, as shown on Supplementary Fig.~\ref{hydrodynamic_model}~(b).

\begin{figure}[h]
\begin{center}
\includegraphics[width=13cm]{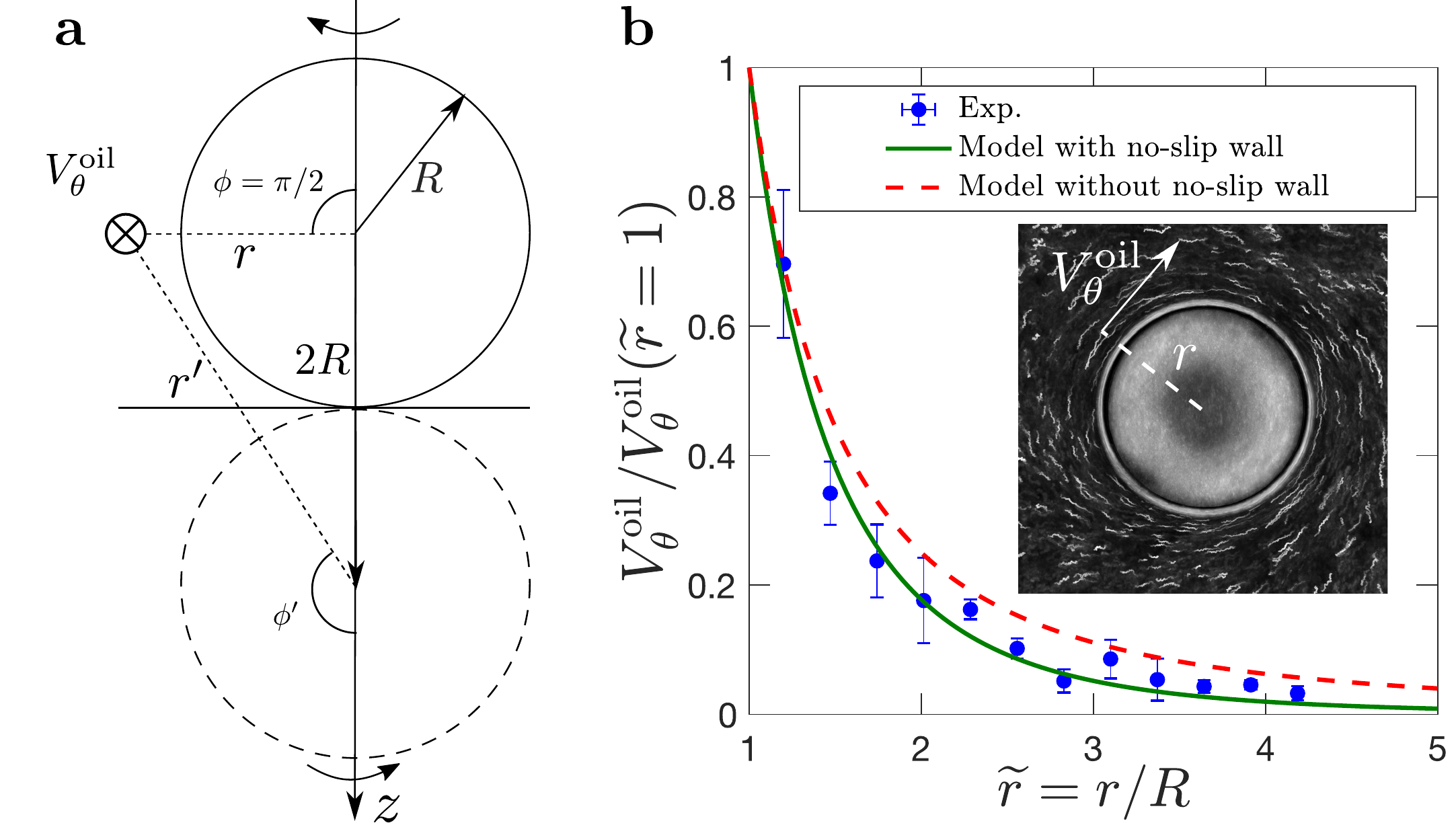}
\end{center}
\caption{\textbf{Hydrodynamic model of a rotating sphere - application to the estimation of the torque exerted by one droplet on the oil.} (a) Definition of the model and hydrodynamic correction due to the presence of the bottom plate of the pool. (b) Test of the hydrodynamic model on experimental data. Dotted line: model without corrective term accounting for the presence of the bottom plate. Solid line: model with corrective term. \hfill \,}
\label{hydrodynamic_model}
\end{figure}

\subsection{Supplementary Note 2. Alignment of magnetotactic bacteria with surfaces under constant magnetic field}

\begin{figure}[h]
\begin{center}
\includegraphics[width=13cm]{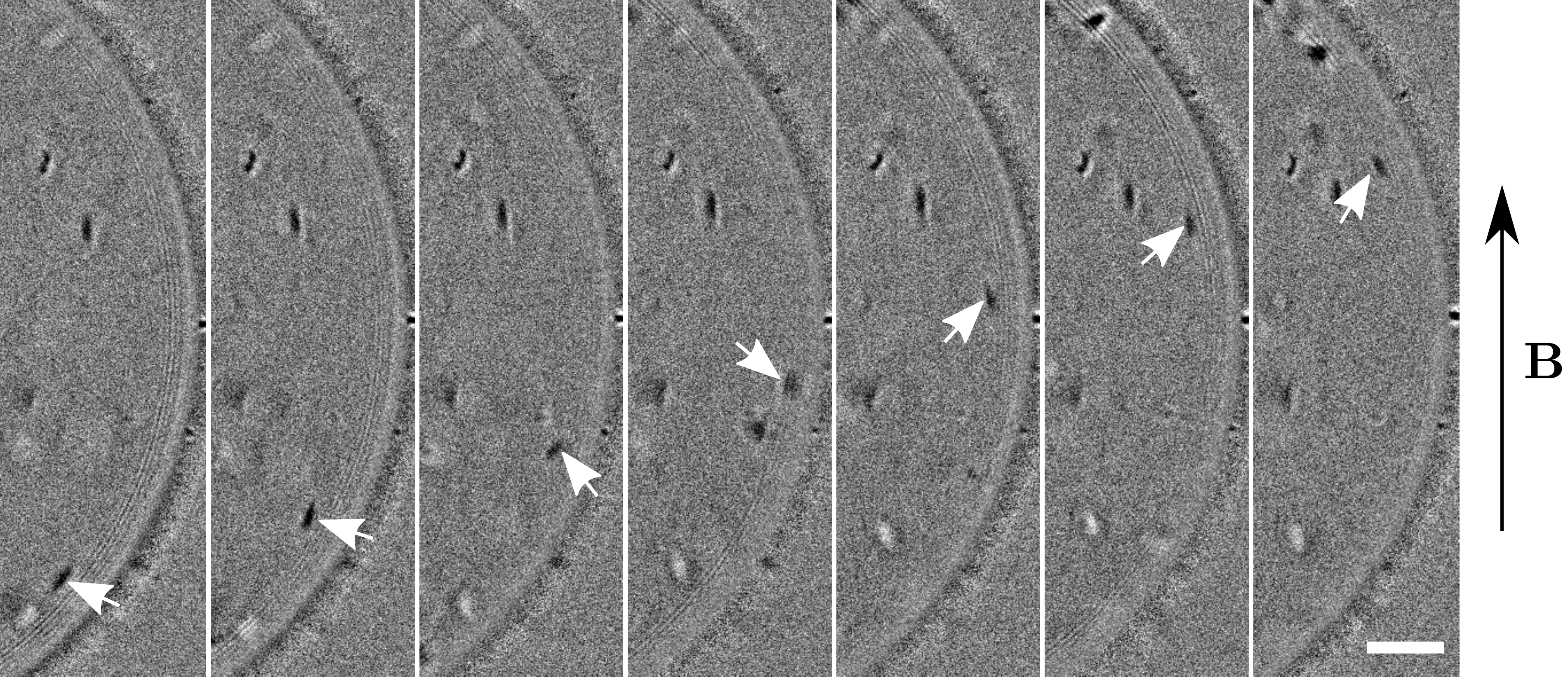}
\end{center}
\caption{\textbf{Image sequence of bacteria motion along a droplet interface.} A bacterium moves along a droplet interface while a magnetic field of \SI{2}{\milli\tesla} is applied. The white scale bar is \SI{10}{\micro\meter} and images are separated by \SI{0.2}{\second}. In average, the bacterium moves with its body aligned with the droplet boundary. The Supplementary Movie 8 shows details about this temporal sequence: the bacterium exhibits a wobbling motion, aligning its body along the magnetic field direction and aligning with the local droplet tangent sequentially.\hfill \,}
\label{alignment_surface}
\end{figure}

We observed single bacteria inside droplets for dilute bacterial suspensions. We observed that bacteria align partially with the droplet boundary in the presence of a magnetic field, wobbling between the magnetic field direction and the local droplet interface. An image sequence (see Supplementary Movie 8) illustrates this particular motion and Supplementary Fig.~\ref{alignment_surface} shows the average motion of one bacterium along the droplet boundary.

\subsection{Supplementary Note 3. MTB velocity distribution and emulsion preparation}

The MTB are grown and prepared as indicated in the Methods part of the article. Here are some details of the bacteria motion characteristics and of the emulsion preparation. 

Before producing the emulsion, bacteria are always harvested after their growth to an optical density OD$=0.12\pm 0.02$. Their motility are always checked quantitatively by performing tracking of single bacteria at $\times 40$ magnification. The velocity distribution (see Supplementary Fig.~\ref{vel_distrib}) is bimodal with two peaks located at 20 and 40 \SI{}{\micro\meter.\second^{-1}}, reminiscent from their reversal motion sequence. This distribution is quantitatively reproducible using our growth protocol.

\begin{figure}[h]
\begin{minipage}[t]{0.30\linewidth}
\centering
\includegraphics[width=5cm]{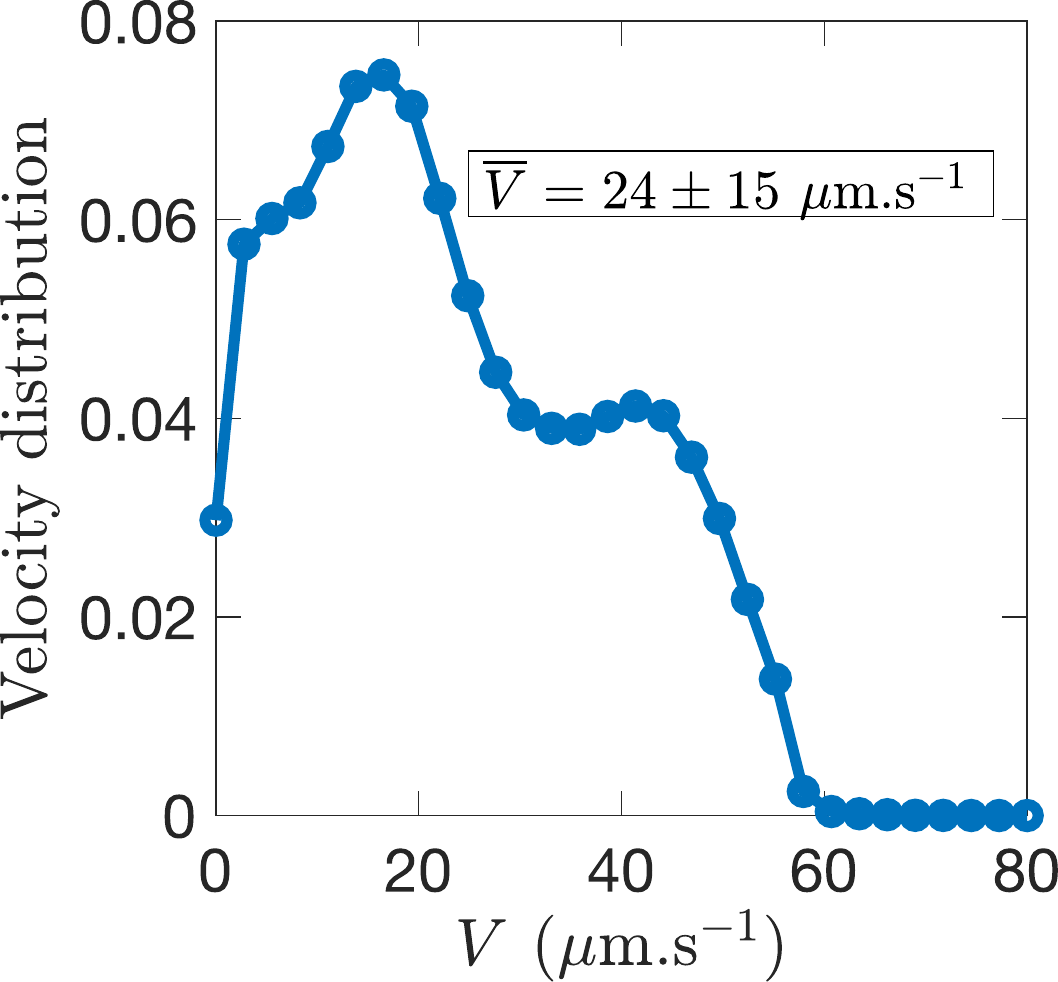}
\caption{Velocity distribution of the MTB after growth (these data are an average for a series of 6 growing sequences corresponding to $\sim$6,000 tracked bacteria)\hfill \,.} 
\label{vel_distrib}
\end{minipage}
\hfill
\begin{minipage}[t]{0.65\linewidth}
\centering
\includegraphics[width=11cm]{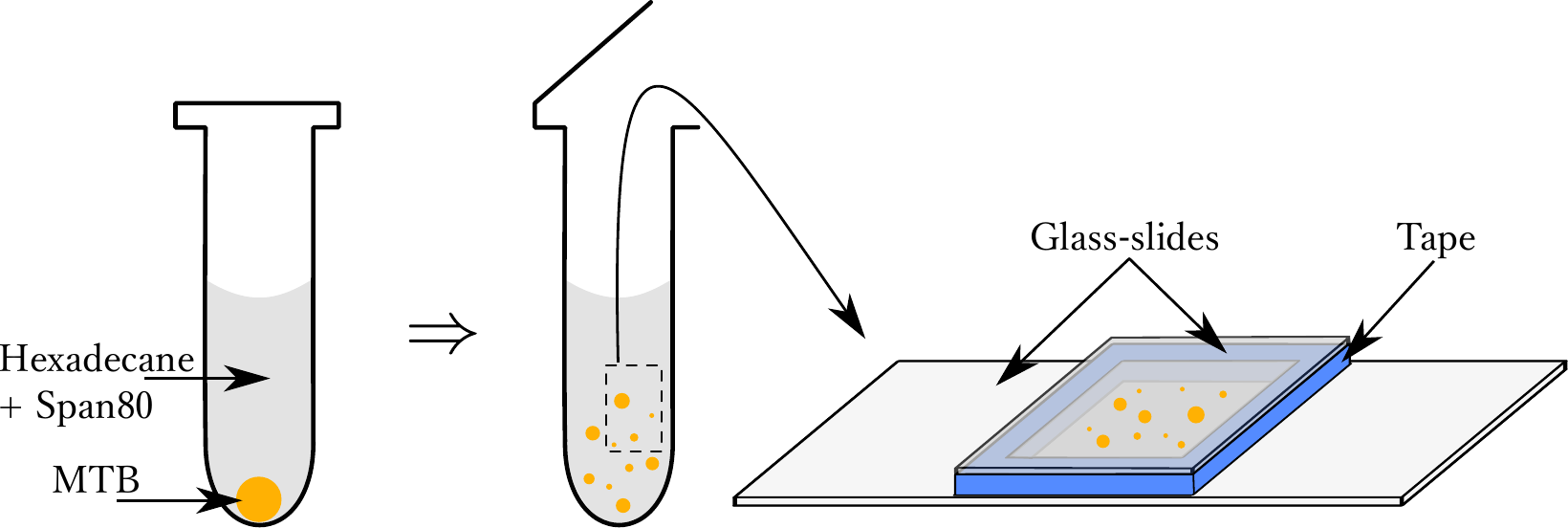}
\caption{Emulsion preparation protocol. We first stir hexadecane oil and MTB suspension. Then, we harvest a sample of the emulsion and transfer it into the experimental pool composed of two glass-slides and a GeneFrame tape.\hfill \,}
\label{emulsion_preparation}
\end{minipage}
\end{figure}

Concerning the emulsion preparation, we fill one \SI{1.5}{\milli\liter}-eppendorf with \SI{500}{\micro\liter} of hexadecane oil. A small amount of bacteria suspension (typically \SI{10}{\micro\liter}) is added to the oil. Then, the eppendorf is gently shaked and a part of the emulsion (\SI{65}{\micro\liter}) is rapidly extracted and placed inside a pool composed by a double sided tape and a cover-glass. The pool is then sealed with a top cover-glass in such a way to avoid bubble formations.

\subsection{Supplementary Note 4. Proof that the total circulation produced by a torque-free swimmer vanishes: detailed version}

A torque and force free swimmer, produces on the far field (distances larger than the body length) a flow  that is well described by that of a force dipole. For a swimmer located at $\mathbf{r}_0$, with director $\hat{\mathbf{n}}$ and force dipole intensity  $p$, the produced velocity field is
\begin{equation}
u_i (\mathbf r) = -\frac{p}{8\pi \eta} \frac{x_i}{|\mathbf x|^3} \left(\delta_{jk}-\frac{x_j x_k}{|\mathbf x|^2}\right) n_j n_k,
\end{equation}
where $\eta$ is the fluid viscosity, $\mathbf x=\mathbf r -\mathbf r_0$, $\delta_{ik}$ is the Kronecker delta, and Einstein notation has been used for repeated indices.
Note that the velocity field is radial ($\mathbf u /\mkern-5mu/ \mathbf x$) with and intensity that depends on the angle. The circulation on a path $\gamma$, which is entirely contained in a plane oriented in the $\hat{\mathbf z}$ direction (see Supplementary Fig.~\ref{fig.scheme}), is 
\begin{equation}
\Gamma_{z,\gamma} = \oint_{\gamma} \mathbf u\cdot d\mathbf r = \int_{A_\gamma} dA\, \omega_z,
\end{equation}
where we have used Stokes theorem, $A_\gamma$ is the area enclosed by the path, and $\mathbf\omega$ is the vorticity field
\begin{equation}
\mathbf \omega =\nabla\times\mathbf u= \frac{3p}{4\pi \eta} \frac{(\hat{\mathbf n}\times\mathbf{x})(\hat{\mathbf n}\cdot\mathbf{x})}{|\mathbf x|^5}.
\end{equation}
In general, $\Gamma_{z,\gamma}$ will be different from zero (see Supplementary Fig.~\ref{fig.circulation.planes}), but here we will show that if the paths are organized in parallel circles around a sphere of radius $R$, the total circulation vanishes. 

\begin{figure}[htb]
\includegraphics[width=.8\columnwidth]{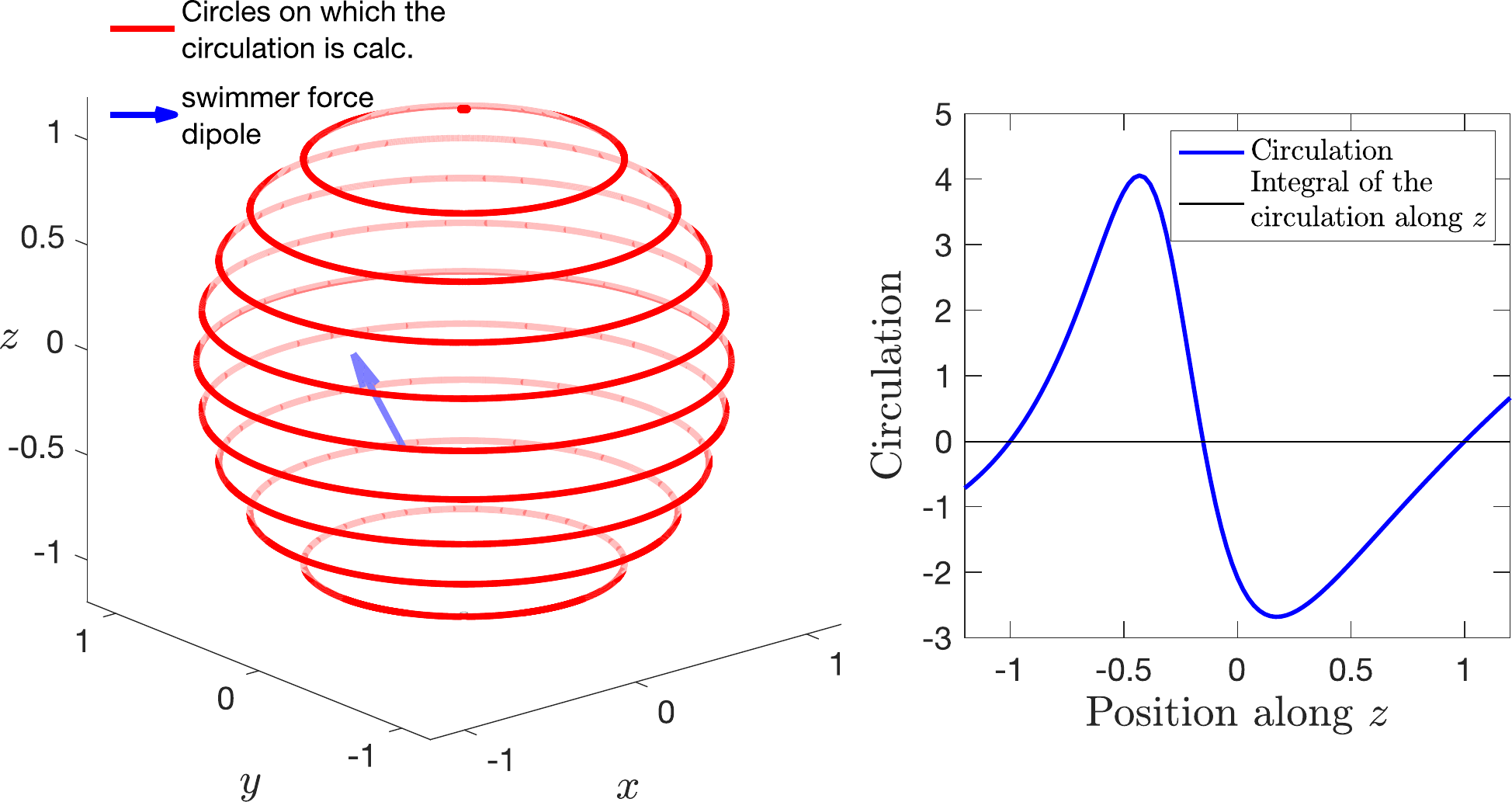}
\caption{Left panel: Representations of a force dipole located in a fixed point inside a sphere and some of the circles used in the circulation calculations. The circles are chosen parallel to the $(x, y)$ plane without loss of generality and are of dimensions such that they altogether enclose a sphere of radius 1.2. The position and the orientation of the swimmer were chosen randomly inside a sphere of radius 1. Right panel: Circulation computed on each of the circles as a function of the vertical $z$ position. The integral of the circulation along $z$ is zero even if the circulation can be locally non-zero. This indicates that the sphere of fluid enclosing the micro-swimmer does not rotate globally, because no net torque is applied on it.\hfill \,  }
\label{fig.circulation.planes}
\end{figure}

First, the total circulation can be written as the integral over the sphere volume of the vorticity. Indeed,
\begin{equation}
\Gamma_z = \int dz\, \Gamma_{z,\gamma}  
 =\int dz \int_{A_\gamma} dA\, \omega_z  = \int_V dV\, \omega_z. \label{eq.Gamma.V}
\end{equation}
It is, then, direct to extend the definition for paths oriented on any direction 
\begin{equation}
\mathbf \Gamma =  \int_V dV \, \mathbf \omega.
\end{equation}

Second, we transform the volume integral into one over the surface of the sphere. For this, we note that
\begin{equation}
\omega_i =  \frac{3p}{4\pi \eta} \epsilon_{ijk} n_j n_m \frac{x_k x_m}{|\mathbf x|^5},\\
= \frac{p}{4\pi \eta} \epsilon_{ijk} n_j n_m \left( \frac{\partial^2}{\partial x_k \partial x_m} \frac{1}{|\mathbf x|} + \frac{\delta_{km}}{|\mathbf x|^3} \right),\\
= \frac{p}{4\pi \eta} \epsilon_{ijk} n_j n_m  \frac{\partial^2}{\partial x_k \partial x_m} \frac{1}{|\mathbf x|},
\end{equation}
where $\epsilon_{ijk}$ is the Levi-Civita tensor and from going form the second to the third line we use that the cross product of equal vectors vanishes. It is possible to use the divergence theorem to obtain
\begin{equation}
\Gamma_i =  \frac{p}{4\pi \eta} \epsilon_{ijk} n_j n_m \int_S \, dS_m \frac{\partial}{\partial x_k} \frac{1}{|\mathbf x|}, \\
=  -\frac{p}{4\pi \eta} \epsilon_{ijk} n_j n_m \int_S  dS_m \, \frac{x_k}{|\mathbf x|^3}. \label{eq.Gamma.Tij}
\end{equation}
The integral $T_{ij}= \int_S dS_j\,  x_i |\mathbf x|^{-3}$ can be evaluated in spherical coordinates. Choosing integration axis such that $\mathbf r_0=z_0\hat{\mathbf z}$, it is direct to verify that $T_{ij}$ is diagonal. Furthermore, the diagonal components are
\begin{equation}
T_{xx}=T_{yy}=R^2\int_0^\textsl{•}\pi d\theta\, \frac{\pi\sin^3\theta}{(1 + z_0^2 - 2 z_0 \cos\theta)^{3/2}} = \frac{4\pi R^2}{3},\\
T_{zz}=R^2 \int_0^\pi d\theta\, \frac{2\pi\cos\theta\sin\theta(\cos\theta-z_0)}{(1 + z_0^2 - 2 z_0 \cos\theta)^{3/2}} = \frac{4\pi R^2}{3},
\end{equation}
implying that it is an isotropic tensor, which can be expressed in any axes as $T_{ij}=4\pi R^2\delta_{ij}/3$. Substituting this result in \ref{eq.Gamma.Tij} implies that $\Gamma_i=0$, concluding the proof. 

For a collection of swimmers, the total induced flow is the sum of those produced by each of them. This linearity, valid for low Reynolds flows, implies that even though the circulation in a particular plane can be finite, its integral on $z$ must vanish. As a consequence, for any distribution of torque-free swimmers in the sphere, the circulation must change sign when measured on different $z$ planes.

\begin{figure}[htb]
\includegraphics[width=.3\columnwidth]{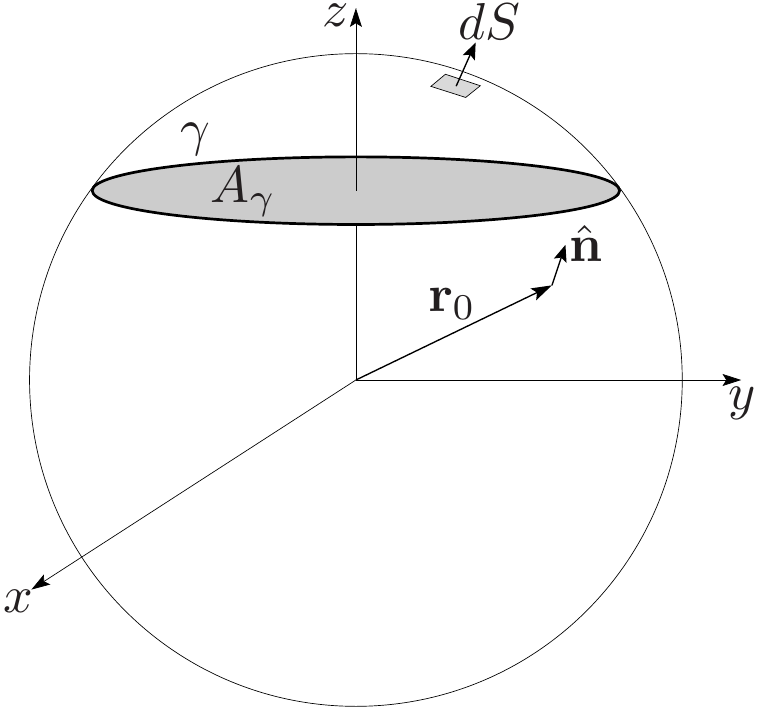}
\caption{Scheme used for the proof. The swimmer is located on any position $\mathbf r_0$ in the interior of the sphere of radius $R$. The path $\gamma$ is a circle, tangent to the sphere, encircling the area $A_\gamma$, which is oriented along the $z$ axis. The total circulation~\ref{eq.Gamma.V} is obtained integrating the circulation $\Gamma_{z,\gamma}$ for all vertical positions of the path, from $-R$ to $R$. The integral in \ref{eq.Gamma.Tij} is done over the surface of the sphere with area elements $dS$. \hfill \,}
\label{fig.scheme}
\end{figure}

\end{document}